\documentstyle[prl,aps]{revtex}
\twocolumn

\begin{document}
\baselineskip=5mm

\title{\large\bf A New Model for Soft Gamma-Ray Repeaters}

\author{K.S. Cheng$^1$ \,\, and \,\, Z.G. Dai$^2$ }

\address{
$^1$Department of Physics, University of Hong Kong, Hong Kong}

\address{
$^2$Department of Astronomy, Nanjing University, Nanjing 210093, 
China}

\address{\mbox{}}
\address{\parbox{14cm}{\rm \mbox{} \mbox{}
We consider a model in which the soft gamma-ray repeaters (SGRs) 
result from young, magnetized strange stars with superconducting cores.
As such a strange star spins down, the quantized 
vortex lines move outward and drag the magnetic field tubes
because of the strong coupling between them. Since the terminations 
of the tubes interact with the stellar crust, the dragged 
tubes can produce sufficient tension to crack the crust. 
Part of the broken platelet will be dragged into the quark core, 
which is only $10^4$ cm from the surface, leading to the deconfinement of 
crustal matter into strange quark matter and thus the release of 
energy. We will show that the burst energy, duration, time interval 
and spectrum for our model are in agreement with the observational results.
The persistent X-ray emission from the SGRs can be well explained by our 
model.
}}
\maketitle

\narrowtext

The soft gamma-ray repeaters (SGRs) are a small, enigmatic class of 
gamma-ray transient sources. There are three known SGRs which are 
charaterized by short rise times (as short as 5 ms) and duration
($\sim 50$-150 ms, FWHM, some less than 16 ms), spectra with characteristic
energies of $\sim$ 30-50 keV  and little or no evolution, and stochastic
burst repetition within a timescale of $\sim 1$ month [1].
SGR 0525$-$66, the source of the 1979 March 5 event, appears to be associated
with the N49 supernova remnant (SNR) in the Large Magellanic Cloud and
hence is apparently the most distant known SGR source at $\sim 55$ kpc
from Earth [2]. The second burster,
SGR 1806$-$20, which produced $\sim 110$ observed bursts during a 7-year
span [3] and recently became active again [4],
appears to be coincident with the SNR G10.0$-$0.3 [5], 
confirming an earlier suggestion [6]. 
Thus, this source is at a distance of $\sim 15$ kpc. 
The third burster, SGR 1900+14, is associated with SNR G42.8+0.6 [7],
and its age is $\sim 10^4$ yrs and its distance from
Earth is $\sim 7$ kpc. Accepting these SGR-SNR associations, the burst
peak luminosities can be estimated to be a few orders higher than the
standard Eddington value for a star with the mass of $\sim 1M_\odot$. For
example, SGR 1806$-$20 has produced events that are $\sim 10^4$ times
the Eddington luminosity [8]. In addition to
short bursts of both hard X-rays and soft $\gamma$-rays, the persistent
X-ray emission was also detected from SGRs [5,7,9].
The luminosities of the persistent X-ray sources are 
$\sim 7\times 10^{35}\,\,{\rm ergs}\,{\rm s}^{-1}$ for SGR 0525$-$66,
$\sim 3\times 10^{35}\,\,{\rm ergs}\,{\rm s}^{-1}$ for SGR 1806$-$20, and
$\sim 10^{35}\,\,{\rm ergs}\,{\rm s}^{-1}$ for SGR 1900+14.
These observations show that the repeaters may be young, magnetized 
neutron stars which power the surrounding luminous plerionic nebulae.

There may be three classes of models for explaining the energy source
of SGRs. In the first class of models, SGRs were thought to result 
from accretion of neutron stars (for a brief review see [10]).
Since the highly super-Eddington 
flux requires the accretion inflow and radiation outflow to be channeled 
in different directions so that it makes any accretion model very 
difficult. Second, it was suggested [11] that glitches
of normal pulsars are an energy source of SGRs. However, the current 
models for pulsar glitches [12,13] seem to give glitching intervals and 
durations much larger than those of SGRs. Moreover, no SGR bursts have 
so far been detected from the Crab pulsar. These two facts may disfavor 
the glitch model for SGRs. Third, it was argued [14] that SGRs are magnetars, 
a kind of neutron stars with superstrong magnetic fields of $\ge 5\times 
10^{14}\,{\rm G}$. Although the motivations for this model 
(e.g., rapid spin-down to 8 s period in $10^4$ years) sound attractive,
there may be several unsettled issues [10], e.g., (i) a power output
from such a strong magnetic field may be inconsistent with the plerion 
energy range; (ii) in such a strong field the radiation output is highly 
anisotropic but the observed shape seems to be angle independent. 
In this letter we suggest that SGRs result from young, magnetized strange 
stars with superconducting cores. 

The structure of strange stars has been studied [15]. 
Strange stars near $1.4M_\odot$ have thin crusts with thickness 
of $\sim 10^4\,$cm and mass of $\sim 10^{-5}M_\odot$. 
Some arguments may be unfavorable to the existence of strange stars.
First, most important, the relaxation behavior of glitches of pulsars
which seem to be isolated neutron stars with masses of $\sim 1.4M_\odot$
is well described by the neutron superfluid vortex creep theory [12],
but the current strange star models scarely explain
the observed pulsar glitches [16]. This may also mean that 
at least pulsars in which glitches occur must be neutron stars, not
strange stars. Second, the conversion of a neutron star to a strange star 
requires the formation of a strange matter seed at a density 
(6-9 times the nuclear matter density) much larger than the central denisty
of the $1.4M_\odot$ star with a rather stiff equation of state [17].
This shows that strange stars are not easy to be produced in the universe.

However, it was argued [18] that 
when neutron stars in low-mass X-ray binaries accrete sufficient 
mass, they may convert to strange stars. This mechanism was further suggested 
as a possible origin of cosmological gamma-ray bursts. In this Letter
we suggest that strange stars may also be formed during the core collapse
of massive stars or during the accretion phase of newly born neutron stars.
The birth rate of strange stars due to these processes must be low. This
is beacuse (i) if the rate were high, the number of resulting 
strange stars would be too high to explain the observed glitch phenomena;
(ii) although the current type II supervova models believe that neutron
stars can be produced during the core collapse of massive stars in some
controversial mass range and the evolution of more massive stars can result
in the formation of black holes, these models have neither given the upper 
limit of the masses of massive stars which evolve to neutron stars nor
the lower limit of the masses of massive stars which evolve to black holes.
We conjecture that massive stars in a narrow mass range may finally evolve
to strange stars. There are two cases for this evolution: (i) during 
the core collapse the nucleon matter directly convert into strange matter 
[19], in which case the shock wave
for the supernova can obtain more energy; (ii) the central density of
a newly born neutron star may reach the deconfinement density due to 
hypercritical accretion in a supernova circumstance [20]
and then the whole neutron star may undergo a phase transition 
to a strange star.

After the birth, a strange star must start to cool due to neutrino
emission. As a neutron star does, the strange star core may become 
superconducting when its interior temperature is below the critical  
temperature. Using a relativistic treatment of BCS theory, Bailin \& Love  
[21] suggested that strange matter forms superconducting. They
showed that the pairing of quarks is most likely to occur in both 
{\em ud} and {\em ss} channels. The pairing state of the former is likely in
s-wave and that of the latter is in p-wave. The superconducting 
transition temperature is about 400 keV. Therefore, a strange star
with age older than $10^3$ years after its supernova birth should 
have a core temperature lower than the normal-superconducting temperature
[22]. The quark superconductor is likely 
to be marginally type II with zero temperature critical field
$B_c\sim 10^{16}$-$10^{17}\,{\rm G}$ [21,23]
which sensitively depends on the interactions between quarks.

On the other hand, the existence of quantized vortex lines in the rotating
core of a strange star is unclear. Since different superconducting 
species inside a rotating strange star try to set up different values 
of London fields in order to compensate for the effect of rotation.
Using the Ginzburg-Landau formulism, Chau [23] showed that instead of 
setting a global London field vortex bundles carrying localized 
magnetic fields can be formed. The typical field inside the vortex core
is about $10^{16}$-$10^{17}\,{\rm G}$ (the accurate value depends on 
strong interaction parameters). Using the similar idea proposed for
the interaction between the proton fluxoids and magnetic neutron
vortics in the core of a neutron star [24], 
he argued that the vortex bundle and the flux tubes can 
interpin to each other by interaction of their core magnetic fields.
He estimated that the pinning energy per intersection is
\begin{equation}
E_p\sim 690 N_{\rm flux}^{1/2}\,\,\,\,{\rm MeV}\,,
\end{equation}
where $N_{\rm flux}$ is the number of flux quantum in a flux tube. Such
strong binding between vortex lines and flux tubes implies that
when the vortex lines moving outward due to spinning-down of the star 
will induce the decay of the magnetic field [23]. One of the
important consequences of this coupling effect will be discussed next text.

We now propose a plate tectonic model for strange stars
which is, in principle, similar to that proposed by Ruderman [24]
for neutron stars. As described in last subsection, there might exist
two different types of quantized flux tubes in the core of a strange star.
The first type of flux tubes are formed when the stellar magnetic field
penetrates through the superconducting core. The second type of flux tubes 
(vortex lines) result from the requirement of minimizing the rotating 
energy of the core superfluid. When the star spin down due to magnetic 
dipole radiation, the vortex lines move outward and pull the flux tubes
with them. Inductive currents do not strongly oppose this flux tube motion 
because of current screening by the almost perfectly diamagnetic 
superconducting quarks. However, the terminations of flux tubes are anchored
in the base of highly conducting crystalline stellar crust.
When the stellar spin-down timescale $\tau_s=\Omega/2\dot{\Omega}$ is
shorter than the typical ohmic diffuse timescale,
\begin{equation}
\tau_D\sim \frac{\sigma A}{4\pi c^2} \sim 3\times 10^4 \sigma_{21}R_6^2
\,\,\,\,{\rm yrs}\,,
\end{equation}
where $\sigma$ is the conductivity and $R_6$ is the radius in units of 
$10^6\,{\rm cm}$. The motion of flux tubes is limited by their
terminations in the crust unless the resulting pull on the crust by these
flux tubes exceeds the crustal yield strength, namely,
\begin{equation}
\frac{BB_c}{8\pi}\sin \theta > \mu \theta_s\frac{l}{R}\,,
\end{equation}
where $B$ is the stellar magnetic field, $\theta$ is the angle bewteen 
the stellar magnetic moment and the flux tubes, $\mu$ is the 
shear modulus, $\theta_s$ is the shear angle, and $l$ is the crustal 
thickness. Substituting the typical values of strange star parameters into 
equation (3), we obtain 
\begin{equation}
\sin \theta \sim \theta > \theta_c\equiv 3\times 10^{-6}
B_{c,17}^{-1}B_{12}^{-1}\theta_{s,-3}\mu_{27}l_4R_6^{-1}\,\,\,\,{\rm rad}\,,
\end{equation}
where $B_{c,17}$ is in $10^{17}\,$G, $B$ in $10^{12}\,$G, $\theta_{s,-3}$
in $10^{-3}$, $\mu_{27}$ in $10^{27}\,{\rm dyn}\,{\rm cm}^{-2}$, and
$l_4$ in $10^4\,$cm. 
When $\theta >\theta_c$, the stellar crust will crack and $\theta$ will be
reduced by an amount $\delta\theta\sim {\rm min}(\theta, \Delta l/R)$
($\Delta l$ is the displacement of the crustal plate). In the case of 
neutron stars, Ruderman [25] estimated that $\Delta l\sim 2\times 
10^2\,{\rm cm}$ for the Crab and Vela pulsars. For a strange star
with a much thinner crust than that of a neutron star, we expect that 
$l>\Delta l >2\times 10^2\,{\rm cm}$, which implies $\delta \theta\sim 
\theta$. Since the flux tubes move outward with the same speed as 
the vortex lines which is given by
\begin{equation}
v\sim \frac{R}{\tau_s}=3\times 10^{-6}R_6\tau_{s,4}
^{-1}\,\,\,\,{\rm cm}\,{\rm s}^{-1}\,,
\end{equation}
where $\tau_{s,4}$ is in $10^4$ yrs,
the time interval between two successive cracking events is estimated to be
\begin{equation}
\tau_{\rm int} \sim \frac{R\delta \theta}{v}\sim 10^6\,
B_{c,17}^{-1}B_{12}^{-1}\theta_{s,-3}\mu_{27}l_4R_6^{-1}\tau_{s,4}
\,\,\,\,{\rm s}\,.
\end{equation}
This value is consistent with 
the typical interval timescale of SGRs.

When the crust cracks, a small platelet could be dragged from 
the crust into the strange matter core which is only $10^4$ cm from
the surface. In the following we make an estimate of the timescale
for the platelet motion. The force pulling the craking platelet horizontally
by the flux tubes is 
\begin{equation}
F_p=\frac{BB_c}{8\pi}\theta A_p\,,
\end{equation}
where $A_p$ is the area of the platelet. Thus, the timescale opening a hole
with area $\sim A_p$ is approximated by
\begin{equation}
\tau_{\rm drag}=\left(2l\frac{A_pM_{\rm cr}}{4\pi R^2}\frac{1}{F_p}\right)
^{1/2} \sim 80\left(\frac{M_{{\rm cr},-5}}{\theta_{s,-3}\mu_{27}R_6}\right)
^{1/2}\,\,\,\,{\rm ms}\,,
\end{equation}
where $M_{{\rm cr},-5}$ is the total
mass of the crust in units of $10^{-5}M_\odot$. 
The durations of the SGRs are expected to be of the same order as
this timescale. As normal matter falls into the core continuously, 
the baryons will deconfine into quarks.
Because each baryon can release $\sim (20-30)$ MeV (the accurate value 
is dependent upon the QCD parameters), 
which are a sum of gravitational energy and deconfinement energy, 
the total amount of energy release is estimated as
\begin{equation}
\Delta\! E\sim 3 \times 10^{42}M_{{\rm cr},-5}(A_p/l^2)l_4^2R_6^{-2}
\,\,\,\,{\rm ergs}\,.
\end{equation}
where $M_{{\rm cr},-5}$ is the total
mass of the crust in units of $10^{-5}M_\odot$. 
At least half of this 
amount will be carried away by thermal photons with the typical energy
$kT\sim 30 \,\,{\rm MeV}$. These thermal photons will be released  
continuously in a timescale of $\sim \tau_{\rm drag}$. 
In the presence of a strong magnetic field, the thermal photons will
convert into electron-positron pairs when
\begin{equation}
\frac{E_\gamma}{2mc^2}\frac{B}{B_q}\sin\Phi\sim \frac{1}{15}\,,
\end{equation}
where $E_\gamma$ is the photon energy, 
$B_q=m^2c^3/\hbar e=4.4\times 10^{13}\,{\rm G}$ and $\Phi$ is the angle 
between the photon propagation direction and the direction of the magnetic 
field [27]. The energies of the resulting pairs will
be lost via synchrotron radiation. The characteristic synchrotron energy is
given by 
\begin{equation}
E_{\rm syn}^{(1)}\sim \frac{3}{2}\gamma_e^2\hbar \frac{eB}{mc}\sin\Phi
\sim 1.5 \,\,\,\,{\rm MeV}\,,
\end{equation}
where $\gamma_e$ is the Lorentz factor of the electron ($\sim 30$).
The first generation of synchrotron photons will be  converted 
into the secondary pairs because the optical depth of photon-photon 
pair production is much large than 1. The characteristic synchrotron 
energy of the secondary pairs is given by
\begin{equation}
E_{\rm syn}^{(2)}\sim \frac{3}{2}\left(\frac{E_{\rm syn}^{(1)}}{2mc^2}\right)
^2\hbar \left(\frac{eB}{mc}\right)\sim 37 B_{12}\,\,\,\,{\rm keV}\,.
\end{equation}
Since the optical depth of photon-electron scattering near the star is 
also much larger than 1,
a radiation-pair fireball in thermal equilibrium will have an initial 
temperature of the same order as $E_{\rm syn}^{(2)}$, and 
will expand adiabatically as a fluid. During the expansion the radiation 
energy is converted into a bulk kinetic energy of the outflow. The fireball 
will cool with $T=T_0(R_0/R)$, and the relativistic Lorentz factor
$\Gamma$ of the bulk motion is $\Gamma=T_0/T=R/R_0$ [28].
Therefore, when the optical depth of the fireball is one, an observer at
infinity will see a blueshifted spectrum with the typical energy of
$\sim E_{\rm syn}^{(2)}$ due to the relativistic bulk motion of the
fireball.
 
Finally, we want to discuss an astrophysical implication of our model.
The persistent X-ray emission from the SGRs was detected. 
If the sources are normal neutron stars with typical magnetic fields of
$\sim 10^{12}\,$G, it is obvious that the persistent X-ray luminosities
from the SGRs may not be explained by the surface blackbody radiation. 
This is because calculations for the cooling of neutron stars [29] predict 
that after $(0.5-1)\times 10^4\,$yrs the bolometric luminosities will be 
at least two orders smaller than the persistent X-ray ones from the SGRs.
Recently Usov [30] suggested that if the sources of the SGRs are magnetars
the persistent X-ray emission may be the thermal radiation of these stars 
which is enhanced by a factor of 10 or more due to the effect of ultrastrong
magnetic fields. We can also explain the observed persistent X-ray emission
by using our model. After each cracking event, at least half of the resulting thermal
energy from the deconfinement of normal matter into strange quark matter
will be absorbed by the stellar core and thus the surface
radiation luminosity at thermal equilibrium may be estimated to be
\begin{eqnarray}
L_x\sim \frac{\xi \Delta\! E}{\tau_{\rm int}}\sim
3 \times 10^{36}\,\xi M_{{\rm cr},-5}(A_p/l^2)l_4R_6^{-1} \nonumber \\
\times B_{c,17}B_{12}\theta_{s,-3}^{-1}\mu_{27}^{-1}\tau_{s,4}^{-1}
\,\,\,\,{\rm ergs}\,{\rm s}^{-1}\,,
\end{eqnarray}
where $\xi$ is a parameter which accounts for both the ratio of the absorbed
thermal energy to the released total energy during a cracking event and the 
ratio of the surface blackbody radiation energy to the absorbed thermal 
energy. We expect that this parameter is of the order of 0.5. Taking 
$B_{c,17}^{-1}B_{12}^{-1}\theta_{s,-3}\mu_{27}\sim 3$ to 
account for $\tau_{\rm int} \sim 3\times 10^6$ s, we have $L_x\sim 5\times 
10^{35}\,\,{\rm erg}\,{\rm s}^{-1}$. This estimated luminosity seems 
to agree with the observed ones from the SGRs.

This work was supported in part by a RGC grant of Hong Kong and 
in part by the National Natural Science Foundation of China.

\begin{center}
------------------------------------
\end{center}

\begin{description}
\item $[1]$ \,J.P. Norris, P. Hertz, K.S. Wood, and C. Kouveliotou,  
              Astrophys. J. {\bf 366}, 240 (1991).
\item $[2]$ \,W.D. Evans {\em et al.}, Astrophys. J. {\bf 237}, L7 (1980);
              T.L. Cline {\em et al.}, {\em ibid.} {\bf 255}, L45 (1982).
\item $[3]$ \,J.G. Laros {\em et al.}, Astrophys. J. {\bf 320}, L111 (1987).
\item $[4]$ \,C. Kouveliotou {\em et al.}, Nature (London) {\bf 368}, 125 
              (1994).
\item $[5]$ \,T. Murakami {\em et al.}, Nature (London) {\bf 368}, 127 (1994).
\item $[6]$ \,S.R. Kulkarni, and D.A. Frail, Nature (London) {\bf 365}, 33
              (1993).
\item $[7]$ \,G. Vasisht, S.R. Kulkarni, D.A. Frail, and J. Greiner,  
              Astrophys. J. {\bf 431}, L35 (1994).
\item $[8]$ \,E.E. Fenimore, J.G. Laros, and A. Ulmer, Astrophys. J.
              {\bf 432}, 742 (1994).
\item $[9]$ \,R.E. Rothschild, S.R. Kulkarni, and R.E. Ligenfelter, 
              Nature (London) {\bf 368}, 432 (1994).
\item $[10]$ E.P. Liang, Astrophys. Space Sci. {\bf 231}, 69 (1995).
\item $[11]$ F. Melia, and M. Fatuzzo, Astrophys. J. {\bf 438}, 904 (1995);
             M. Fatuzzo, and F. Melia, {\em ibid.} {\bf 464}, 316 (1996).
\item $[12]$ D. Pines, and M.A. Alpar, Nature (London) {\bf 316}, 27 (1985).
\item $[13]$ B. Link, and R.I. Epstein, Astrophys. J. {\bf 457}, 844 (1996).
\item $[14]$ R.C. Duncan, and C. Thompson, Astrophys. J. {\bf 392}, L9 (1992);
             C. Thompson, and R.C. Duncan, Mon. Not. R. Astron. Soc. 
             {\bf 275}, 255 (1995).
\item $[15]$ C. Alcock, E. Farhi, and A. Olinto, Astrophys. J. {\bf 310}, 
             261 (1986); P. Haensel, J.L. Zdunik, and R. Schaeffer, 
             Astron. Astrophys. {\bf 160}, 121 (1986).
\item $[16]$ M.A. Alpar, Phys. Rev. Lett. {\bf 58}, 2152 (1987).
\item $[17]$ G. Baym, in {\em Neutron Stars: Theory and Observations} eds. 
             J. Ventura and D. Pines (Kluwer Academic Publishers), 21 (1991).
\item $[18]$ K.S. Cheng, and Z.G. Dai, Phys. Rev. Lett. {\bf 77}, 1210 (1996).
\item $[19]$ N.A. Gentile {\em et al.}, Astrophys. J. {\bf 414}, 701 (1993);
             Z.G. Dai, Q.H. Peng, and T. Lu, {\em ibid.} {\bf 440}, 815 (1995).
\item $[20]$ R.A. Chevalier, Astrophys. J. {\bf 346}, 847 (1989);
             G.E. Brown, {\em ibid.} {\bf 440}, 270 (1995).
\item $[21]$ B. Bailin, and A. Love, Phys. Rep. {\bf 107}, 325 (1984).
\item $[22]$ O.G. Benvenuto, H. Vucetich, and J.E. Horvath, Nucl. Phys. B 
             (Proc. Suppl.) {\bf B24}, 160 (1991).
\item $[23]$ H.F. Chau, Astrophys. J., in press (1997).
\item $[24]$ H.F. Chau, K.S. Cheng, and K.Y. Ding, Astrophys. J. {\bf 399}, 
             213 (1992).
\item $[25]$ M.A. Ruderman, Astrophys. J. {\bf 366}, 261 (1991).
\item $[26]$ M.A. Ruderman, preprint (1996).
\item $[27]$ M.A. Ruderman, and P.G. Sutherland, Astrophys. J. {\bf 196}, 
             51 (1975).
\item $[28]$ J. Goodman, Astrophys. J. {\bf 308}, L47 (1986);
             A. Shemi, and T. Piran, {\em ibid.} {\bf 365}, L55 (1990).
\item $[29]$ K. Nomoto, and S. Tsuruta, Astrophys. J. {\bf 312}, 711 (1987).
\item $[30]$ V.V. Usov, Astron. Astrophys. {\bf 317}, L87 (1997).
\end{description}

\end{document}